\documentclass[conference, compsoc]{IEEEtran}

\usepackage{endnotes}
\usepackage{siunitx}
\usepackage{cite}
\usepackage{graphicx}
\usepackage{textcomp}
\usepackage{listings}
\usepackage{url}
\usepackage{array}
\usepackage{xspace}
\usepackage[table,xcdraw]{xcolor}

\usepackage{enumitem}

\definecolor{codegreen}{rgb}{0,0.6,0}
\definecolor{codegray}{rgb}{0.5,0.5,0.5}
\definecolor{codepurple}{rgb}{0.58,0,0.82}
\definecolor{backcolour}{rgb}{0.95,0.95,0.92}

\lstdefinestyle{mystyle}{
    backgroundcolor=\color{backcolour},   
    commentstyle=\color{codegreen},
    keywordstyle=\color{magenta},
    numberstyle=\tiny\color{codegray},
    stringstyle=\color{codepurple},
    basicstyle=\ttfamily\footnotesize,
    breakatwhitespace=false,         
    breaklines=true,                 
    captionpos=b,                    
    keepspaces=true,                 
    numbers=left,                    
    numbersep=5pt,                  
    showspaces=false,                
    showstringspaces=false,
    showtabs=false,                  
    tabsize=2
}

\begin{document}

\title{SoK: A Beginner-Friendly Introduction to Fault Injection Attacks}

\author{Anonymous Author(s)}

\author{
    \IEEEauthorblockN{Christopher Simon Liu\IEEEauthorrefmark{1}}
	\IEEEauthorblockA{The Ohio State University\\
		liu.8689@osu.edu}
	\and
	\IEEEauthorblockN{Fan Wang\IEEEauthorrefmark{1}}
	\IEEEauthorblockA{The Ohio State University\\
		wang.13244@.osu.edu}
	\and
    \IEEEauthorblockN{Patrick Gould}
	\IEEEauthorblockA{The Ohio State University\\
		gould.279@osu.edu}
	\and
	\IEEEauthorblockN{Carter Yagemann}
	\IEEEauthorblockA{The Ohio State University\\
		yagemann.1@osu.edu}
    
}	

\maketitle

\begingroup\renewcommand\thefootnote{\IEEEauthorrefmark{1}}
\footnotetext{The two lead authors contributed equally to this work.}
\endgroup

\thispagestyle{plain}
\pagestyle{plain}

\begin{abstract}
Fault Injection is the study of observing how systems behave under unusual stress, environmental or otherwise. In practice, fault injection involves testing the limits of computer systems and finding novel ways to potentially break cyber-physical security.

The contributions of this paper are three-fold. First, we provide a beginner-friendly introduction to this research topic and an in-depth taxonomy of fault injection techniques. Second, we highlight the current state-of-the-art and provide a cost-benefit analysis of each attack method. Third, for those interested in doing fault injection research, we provide a replication analysis of an existing vulnerability detection tool and identify a research focus for future work.
\end{abstract}


%
\IEEEpeerreviewmaketitle

\section{Introduction}
\label{s:introduction}
The most publicly known example of fault injection in computer systems would be in 2013 when an ionizing particle caused a Super Mario ``speed runner" to complete in record time~\cite{supermario}. A more serious example is an incident in 2004 when a cosmic ray caused an election voting system to count 4096 extra votes~\cite{evoting}. Needless to say, fault injections lead to unintended code values and unintended code executions. If these faults are intentional, they are a significant cyber threat; Fault injections are active physical attacks.

The oldest research paper on the topic of fault injection was in 1975, observing how cosmic rays cause unexpected triggering of digital circuits in satellites~\cite{binder1975satellite}. In 1997, a theoretical model demonstrated that it was possible to compromise encrypted systems with hardware faults~\cite{boneh1997importance}. This seminal work sparked a substantial body of research focused on exploiting hardware vulnerabilities.

Since then, many fault injection techniques have been created, researched, and published. Of the various injection techniques, voltage glitching ~\cite{11zussa2012investigation, 12zussa2014analysis, 13selmane2011security, 14kim2014flipping}, clock glitching ~\cite{15obermaier2017fuzzy,16korak2014effects}, laser fault
injection ~\cite{17selmke2016precise, 18dutertre2018laser}, and electromagnetic fault injection ~\cite{19dehbaoui2012injection, 20dehbaoui2012electromagnetic} are among the most prevalent.
 
One way to describe fault injection is to compare it with ``torturing" computer chips until they submit or reveal secrets~\cite{securityBSidesLondon}. From the perspective of an attacker, fault injection techniques are dangerously useful for bypassing secure boot~\cite{bypassingSB}, enabling arbitrary code execution~\cite{switch}, escalating privileges~\cite{ps3}, and extracting keys~\cite{aesdfa}. 


When studying fault injections, we separate the process into two stages: Fault Modeling (examining attack methods) and Fault Analysis (examining faulty outcomes). The following is a list of fault models used in the literature, providing an overview of how fault injection is used.

\begin{description}[style=unboxed]
  
\item[Bit flip] 
    Reverse Engineering Neural Networks~\cite{reverseEngineeringNN}, 
    Adversarial Neural Networks~\cite{adversarialNN}
    
\item[Bit set/reset] 
    Breaking Substitution-Permutation Network (SPN) Block Ciphers~\cite{breakingSPN}
    
\item[Random byte] 
    Breaking Encryption with Differential Fault Analysis (DFA)~\cite{breakingDFA}
    
\item[Instruction skip] 
    Escalating Privileges~\cite{privilegeEscalation},
    Extracting Keys~\cite{simpleKeyExtraction},
    Neural Network Misclassifications~\cite{neuralNetworkMisclassification}
    
\item[Execution faults] 
    Breaking Physically Unclonable Functions (PUFs)~\cite{breakingPhysicallyUnclonable}
    
\item[Stuck-at faults] 
    Breaking Symmetric Cryptography~\cite{breakingSymmetricCryptography},
    Breaking Random Number Generators (RNG)~\cite{breakingRNG}
    
\end{description}

In response to Fault Injection Attacks (FIA), various hardware-based tools ~\cite{21richter2021fiver, 22arribas2020cryptographic, 23wang2021sofi, 24bertoni2003error} and software-based tools ~\cite{25potet2014lazart, 26lacombe2024combining, 27brejon2019fault, 28mahmoud2019minotaur, 29given2017automated, 30riviere2014combining, 31dureuil2015code} were created to assess device vulnerability. 

In this paper, we focus on the software-level manifestation of injected faults. We complete an examination and replication analysis of an existing software-based FIA vulnerability detection tool. We perform a literature review to compare varying fault models. We identify limitations in these existing fault models. We then highlight an area for further research. 


Despite FIAs involving both hardware (for inducing) and software (for exploiting), existing work focuses exclusively on one or the other. We hypothesize that the identified limitations are addressed by integrating both, bridging the gap in FIA modeling.

In summary, we make the following contributions.

\begin{itemize}
    \item We provide a systematic review of existing work.
    \item From our review, we identify a critical gap in fault modeling between hardware and software.
    \item We propose a synthesis solution to identified limitations to motivate future work.
\end{itemize} 



\clearpage

\section{Foundations}
\label{s:foundations}
This section provides the essential knowledge for understanding FIAs. 
We show Metastability's role in device vulnerability and how it leads to FIA through Single Event Upsets and Transient Effects.

\subsection{Metastability}

\begin{figure}
    \centering
    \includegraphics[width=1.0\linewidth]{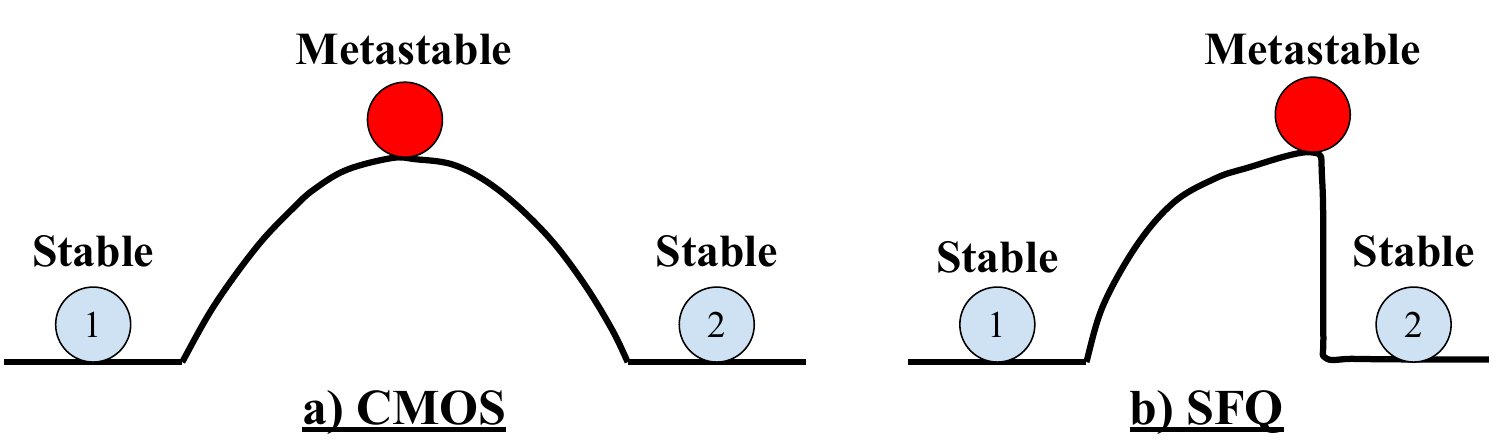}
    \caption{Mechanical Metastability: The Ball on a Hill in~\cite{metastabilityfigure}}
    \label{fig:metastabilityfigure}
\end{figure}

Metastability is an unstable state in electronics that leads to system failure when it occurs in integrated circuits (ICs). A useful metaphor (Figure~\ref{fig:metastabilityfigure}) for this phenomenon is a ball positioned at the top of a hill; it may eventually fall into one of two adjacent valleys, but it is uncertain which valley it will end up in. Similarly, every bistable storage element in an IC may resolve to a `0' or `1'. The metastability of a storage element is induced by a voltage level that lies between the logical states of `0' and `1'.

Faulty outputs caused by metastability propagate through ICs, leading to more significant errors. However, it is impossible for a digital circuit to completely detect or prevent metastability~\cite{marino1981general}. Fault injection techniques, such as voltage and clock glitches, increase the probability of metastability occurring.

\subsection{Single Event Upset}

\begin{figure}
    \centering
    \includegraphics[width=1.0\linewidth]{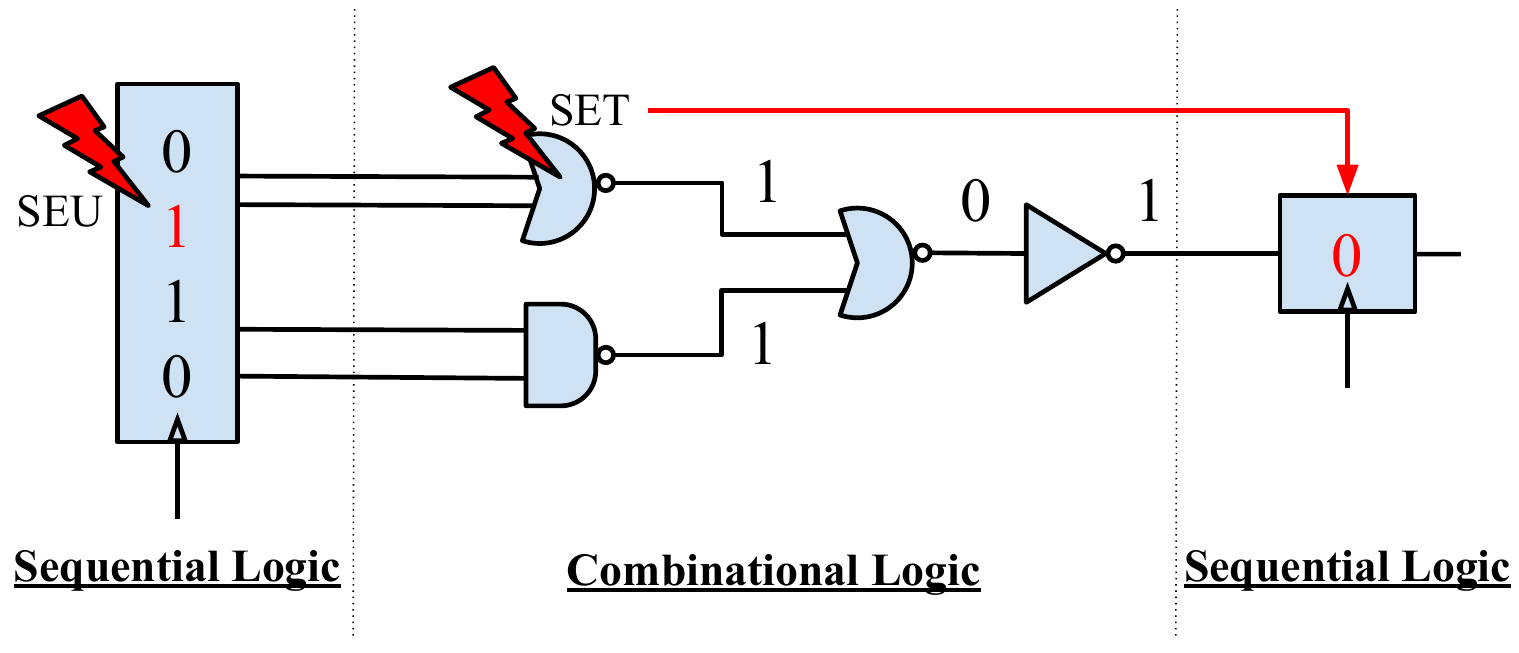}
    \caption{Single Event Upset and Single Event Transient Effects in~\cite{circuitfigure}}
    \label{fig:circuitfigure}
\end{figure}

Single Event Upsets (SEU) are changes in the state of a transistor, altering the logic of a digital circuit. This leads to a Single Event Transient (SET) effect where an incorrect result is interpreted as the final value, leading to the propagation of errors and a fault injection (Figure~\ref{fig:circuitfigure}).

As mentioned in the introduction, the first documented occurrence of SEU was reported in 1975, where four upsets were observed during 17 years of satellite operation in space~\cite{binder1975satellite}. As the occurrence of SEUs became more evident, the hardening of integrated circuits (ICs) emerged as a primary mitigation strategy, proposed by several research groups~\cite{Andrews1982Single, Giddings1985Single}. Further research revealed that the ionization processes responsible for SEUs vary~\cite{wyatt1979soft, guenzer1979single}. Today, there is growing interest in understanding the impact of SEUs on machine learning systems, as machine learning has become increasingly prominent~\cite{chen2022solar, yan2020single}.

\section{Attack Taxonomy}
\label{s:attacks}
This section classifies each fault injection technique or ``attack vector". 
We show that the underlying mechanism behind attack methods are ``Timing Constraint Violations" as well as the Precision vs. Cost trade-off of each method. 

\begin{figure}
    \centering
    \includegraphics[width=1.0\linewidth]{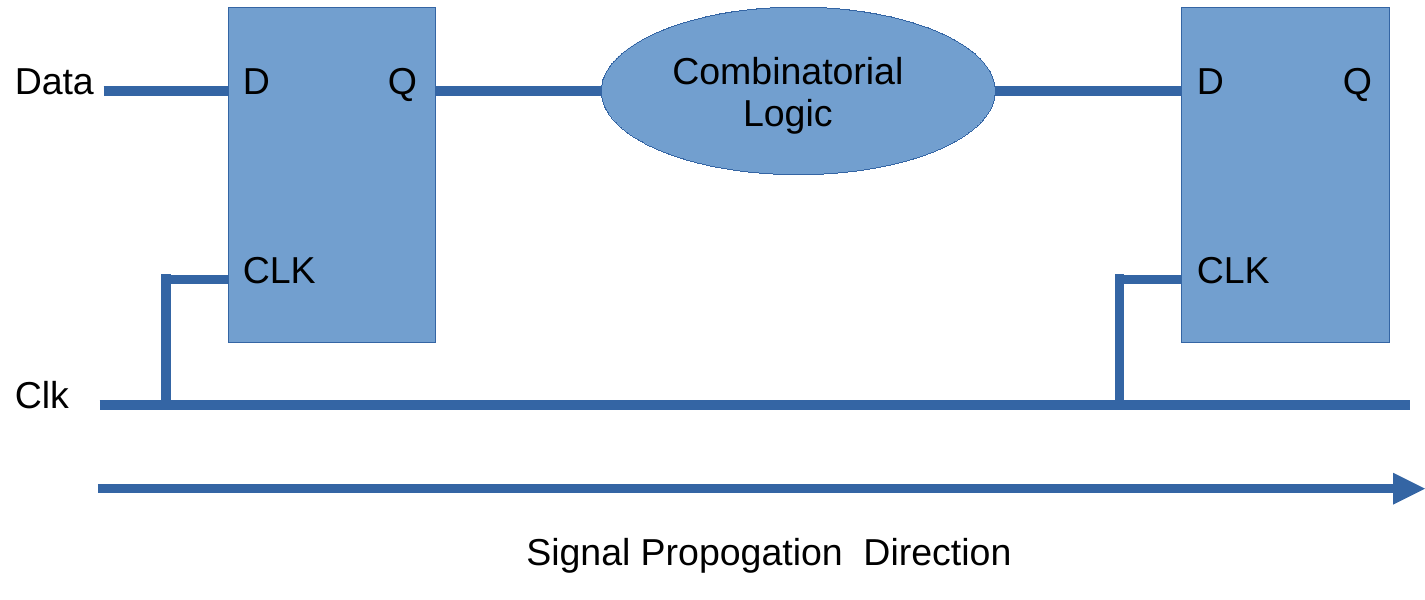}
    \caption{Combinatorial Logic}
    \label{fig:timingconstraints}
\end{figure}

\subsection{Timing Constraint Violations}
Timing constraints are fundamental requirements for the timing of digital ICs that must be met to ensure correct functionality. Most digital ICs are generalized by an architecture in which combinatorial logic is located between two D flip-flops (Figure~\ref{fig:timingconstraints}). The correctness of the data latched into the downstream D flip-flop is critically dependent on avoiding timing violations. Refer to Appendix~\ref{FourthAppendix} for the specific parameters of timing constraints.

\begin{figure}
    \centering
    \includegraphics[width=1.0\linewidth]{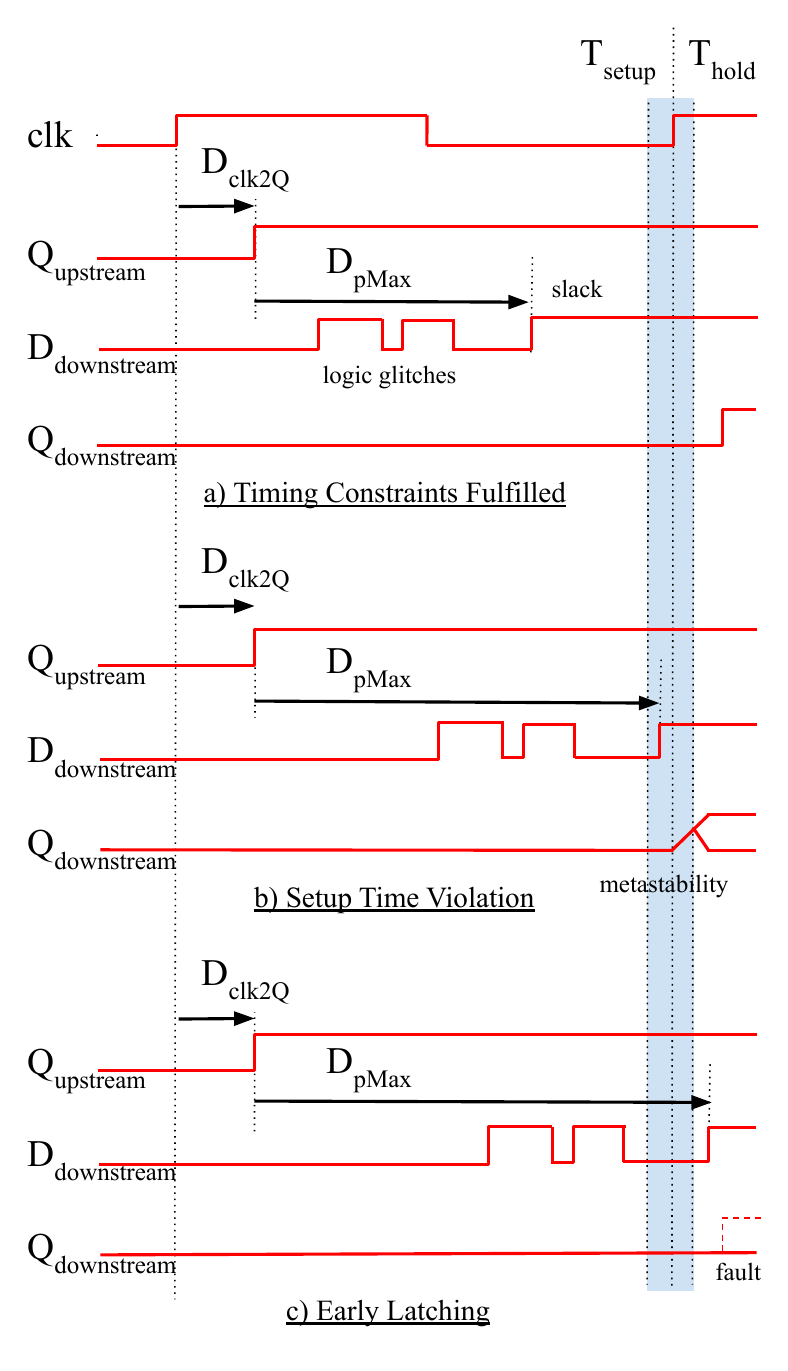}
    \caption{Voltage Attack Waveforms in~\cite{11zussa2012investigation} }
    \label{fig:zussa2012}
\end{figure}

In short, when timing constraints are violated, the setup time requirement will no longer be satisfied (Figure~\ref{fig:zussa2012}). This leads to incorrect values being set, errors propagated, and a Fault Injection Attack on the circuit.

The underlying mechanism behind voltage glitches is hypothesized to be caused by metastability resulting from timing constraint violations~\cite{djellid2006supply}. This hypothesis was substantiated by later experiments, finding that the faults produced by clock glitches and voltage glitches were identical~\cite{zussa2013power}. We therefore lump together Clock and Voltage Glitching because their method is \textit{functionally the same}.

Similarly, the underlying mechanism behind Electromagnetic Fault Injection (EMFI) is theorized to be the violation of timing constraints~\cite{dehbaoui2012electromagnetic}. Another explanation behind EMFI is the disruption of the sampling process of D flip-flops~\cite{ordas2017electromagnetic}. Experiments have shown that both mechanisms may coexist during an EMFI attack~\cite{nabhan2023highlighting}.

Additionally, laser fault injection directly flips individual bits with ionizing radiation. This alters the state of a logic gate, leading to errors in the output~\cite{petersen2011single}.


\subsection{Voltage Glitch}
Voltage glitches involve briefly changing the supplied voltage of a device outside the specified intended range. Imagine switching a device off and on before it can properly react. This is voltage glitching. 

Voltage glitches have two primary effects on digital circuit combinatorial logic. The first effect occurs directly on the victim gate, and the second arises from the propagation of the error through logic~\cite{djellid2006supply}. In this paper, we adopt the same classification to analyze the impact of voltage glitches on Integrated Circuits (IC).

Refer to Appendix~\ref{FifthAppendix} for details of how changes in voltage break timing constraints.

\subsubsection{Underpowering}
As the name suggests, underpowering supply glitches result from lowering the IC's voltage level below the nominal value. Underpowering has been widely utilized in FIAs~\cite{13selmane2011security, 11zussa2012investigation, barenghi2010low, zussa2013power}~\cite{martin2014fault}. Multiple experiments have substantiated underpowering by connecting a device under test (DUT) to a remotely controlled power supply~\cite{13selmane2011security, 11zussa2012investigation}.

\textbf{Considerations:}
It is evident that underpowering is relatively easy to carry out and cost-effective. It also comes with significant drawbacks. It is challenging to precisely inject faults into a DUT with this method, given that the supply voltage affects the entire IC.

\subsubsection{Negative Power Supply Glitch}
Similar to underpowering, negative power supply glitches result from a precise drop in voltage. A negative power supply glitch typically lasts for a shorter duration with a sharp voltage waveform, whereas the waveform of underpowering is flatter~\cite{12zussa2014analysis}. Negative power triggers a glitch at the exact moment when the desired round of computation begins, allowing for more targeted fault injection~\cite{13selmane2011security}. 

\textbf{Considerations:}
It is a more precise form of underpowering. It requires a deeper understanding of the IC and the specific processes running on it, making it more difficult.

\subsubsection{Positive Power Supply Glitch}
At first glance, the concept of a positive power glitch may seem counterintuitive, as an increase in supply voltage typically results in a decrease in data propagation time. However, a positive power glitch induces a damping oscillation in the voltage supply, indirectly causing the voltage to drop below the nominal level, leading to timing constraint violations and faults~\cite{11zussa2012investigation}.

\textbf{Considerations:}
It is an unintended negative power glitch. It is not widely utilized in FIAs.

\subsection{Clock Glitch}
Clock glitches involve injecting short signals that deviate from the nominal clock signal frequency. Clock glitches share the same underlying mechanism as voltage glitches~\cite{zussa2013power}. Since this has already been discussed previously, we will keep our explanation here brief. 

\begin{figure}
    \centering
    \includegraphics[width=1.0\linewidth]{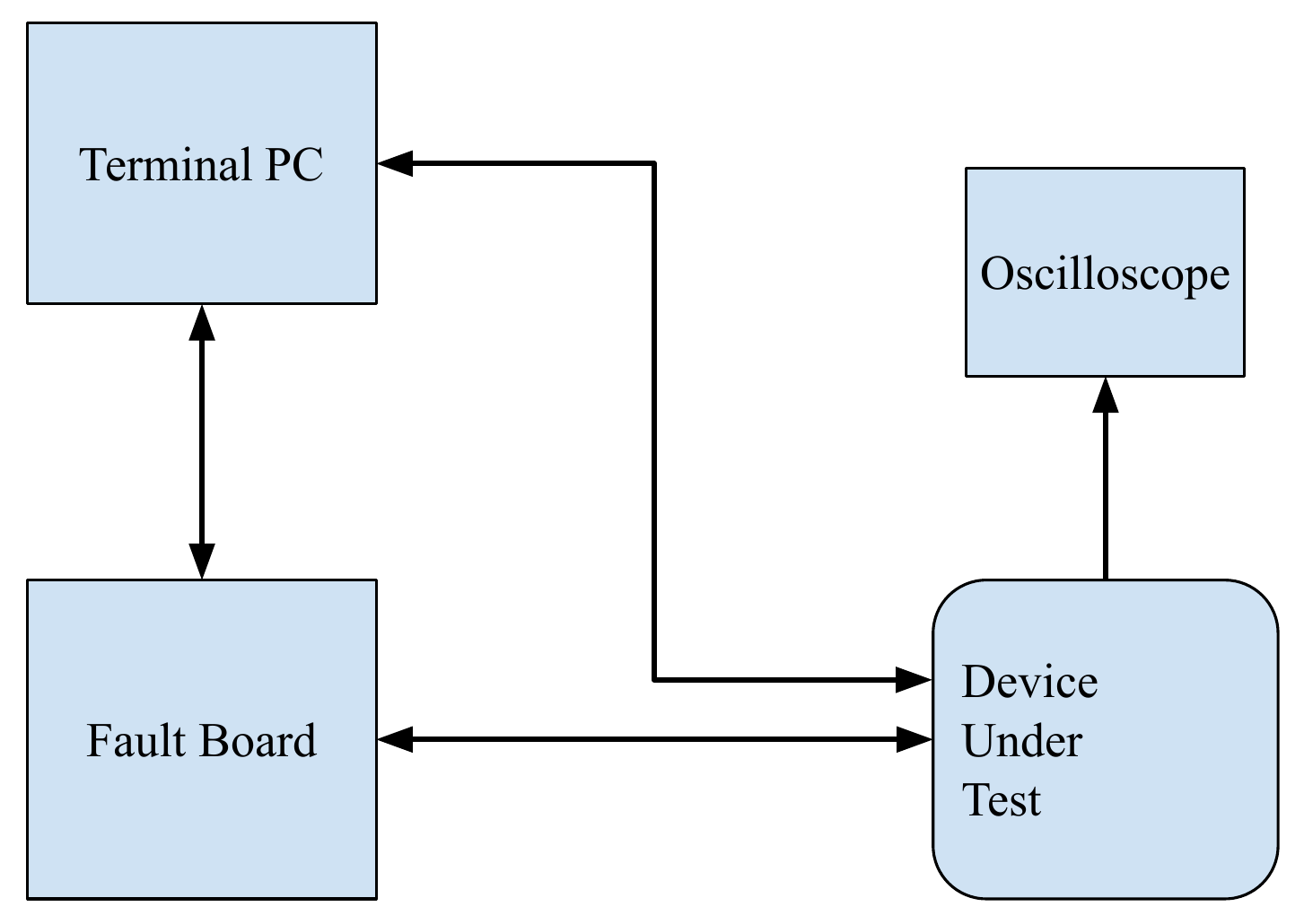}
    \caption{Diagram of the Clock Glitch Environment in~\cite{korak2014clock} }
    \label{fig:korak2014}
\end{figure}

Injecting a faulty clock signal triggers incorrect execution by violating timing constraints. The challenge in this process is how to generate the glitches. The use of precise clock glitches requires several preconditions due to the high precision needed in signal generation. This has led to a divergence in methodologies for generating clock glitches.

\subsubsection{Precisely Timed Glitch}
As the name suggests, this approach depends on generating precise and deterministic clock signals. To satisfy these conditions, the target device should have already been well studied before. Existing experiments demonstrate that precisely timed clock glitches are highly reproducible with 100\% reliability.\cite{16korak2014effects}. Glitches are injected after a predefined number of clock cycles, calculated by a FPGA fault board~\cite{matsubayashi2016clock}. Other experiments, like the one shown in (Figure~\ref{fig:korak2014}), also show the reproducibility of clock glitch FIAs~\cite{korak2014clock}.

\textbf{Considerations:}
It is deterministic (not random), reproducible, and reliable. It is \textit{a lot} of work to precisely time.

\subsubsection{Fuzzy Glitch}
In contrast to precisely timed clock glitches, fuzzy-glitch techniques generate randomized clock signals. A fuzzy clock signal generator, utilizing two ring oscillators operating at different frequencies, produces a fuzzy glitch signal, which is both random and high-frequency~\cite{15obermaier2017fuzzy}. Fuzzy glitches are shown to achieve a 75\% success rate under optimal conditions, with a remaining success rate above 50\% across a wide range of parameters~\cite{15obermaier2017fuzzy}.

\textbf{Considerations:}
It is non-deterministic (random). It is easier than calculating the precise time. Similar to traditional ``fuzzing", it takes a brute-force approach in finding both useful and useless inputs.

\subsection{RowHammer}
This is a voltage glitch in memory instead of the processor. While technically a software attack, the underlying cause of a RowHammer Attack is the leakage of charge in certain DRAM cells, triggered by voltage fluctuations. 

\begin{figure}
    \centering
    \includegraphics[width=1.0\linewidth]{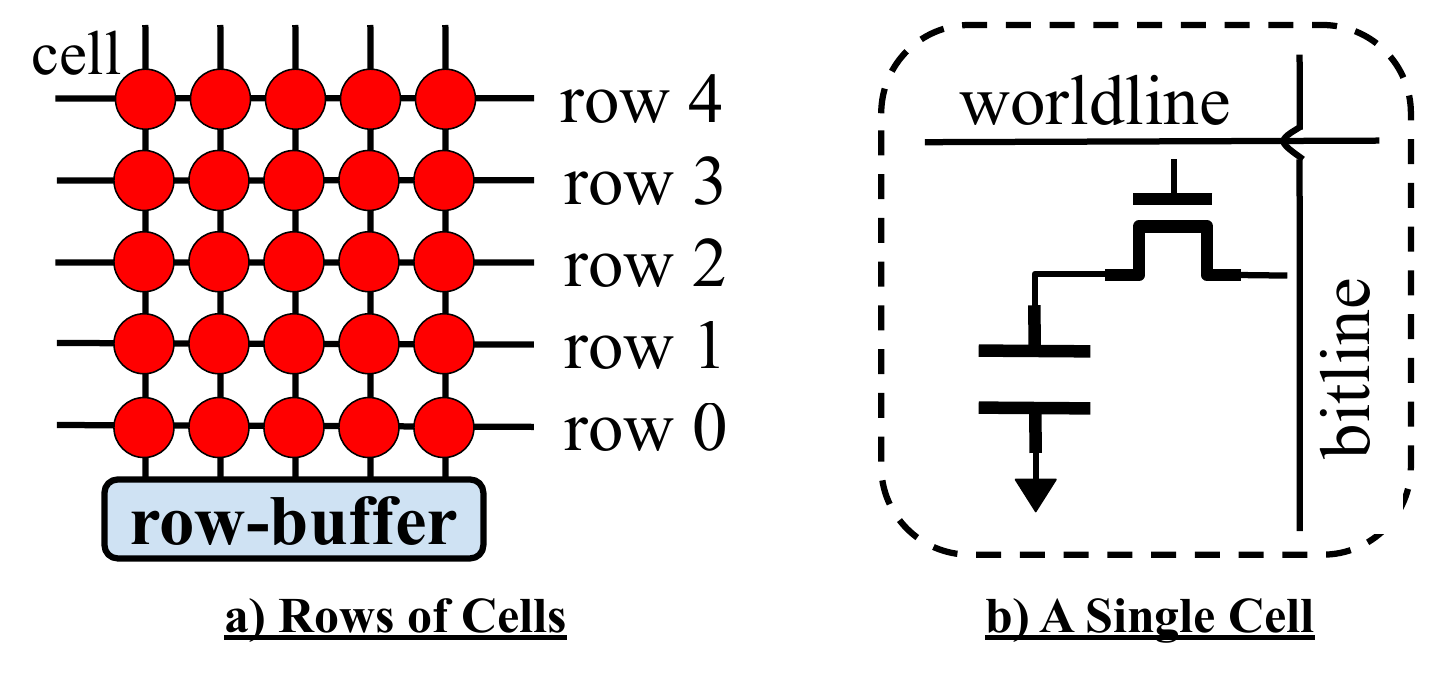}
    \caption{DRAM Organization in~\cite{14kim2014flipping} }
    \label{fig:kim2014}
\end{figure}

RowHammer was first discovered in 2014~\cite{14kim2014flipping}. As technology advanced, the density of DRAM cells has increased, which in turn has made these cells more susceptible to errors due to their close proximity. To execute a RowHammer attack, the wordline must be toggled on and off repeatedly, which is achieved by reading the same DRAM row multiple times. Refer to Figure~\ref{fig:kim2014} for layout. 

Repeatedly toggling the ``wordline" activation wire's voltage causes some cells in neighboring rows to leak charge at an accelerated rate. This leakage prevents the cells from retaining their charge for the standard refresh interval, leading to data loss and disturbance errors~\cite{14kim2014flipping}. Experiments have shown that this is achieved in as few as 139,000 reads to a DRAM address~\cite{14kim2014flipping}.

\textbf{Considerations:}
It is technically a voltage glitch in memory instead of the processor which does not require physical access. It is the only attack method that is completely software-based.
It still has challenges in placing the target data in specific physical memory locations that are vulnerable to fault injection.  

\subsubsection{Deterministic RowHammer}
In 2016, the deterministic RowHammer attack was proposed, targeting ARM/Android systems~\cite{van2016drammer}. This explored the feasibility of using RowHammer in modern always-on commodity mobile systems. In addition, the attack used only standard features, reducing the baseline requirements for a successful attack.

To ensure that sensitive data reside within exploitable memory components, deterministic memory massaging techniques are employed. For example, ``Phys. Feng Shui" is a method of positioning sensitive data in physically vulnerable memory addresses, utilizing predictable memory reuse patterns and occupying unused memory chunks~\cite{van2016drammer}. Other techniques include Memory Deduplication to make the RowHammer attack deterministic~\cite{razavi2016flip}.

\textbf{Considerations:}
It focuses on precise memory locations that are certain to be vulnerable.
 
\subsubsection{Probabilistic RowHammer}
In 2016, it was also found that the use of non-temporal store instructions, where data is created and not immediately consumed, triggers RowHammer vulnerabilities~\cite{qiao2016new}. This means that benign code, which do not follow the normal cache-coherency rules, can sometimes cause a RowHammer event.

To ensure that sensitive data reside within exploitable memory components, probabilistic memory massaging techniques are also employed. Memory Spraying is a method in which the target memory is filled with page tables~\cite{seaborn2015exploiting}. Due to the probabilistic nature of this method, not every mapped page table is physically vulnerable to RowHammer. However, the specific layout of the sprayed physical memory can increase the likelihood that sensitive page tables will be located in a vulnerable position. Other techniques include virtualization to achieve a similar goal~\cite{xiao2016one}.

\textbf{Considerations:}
It focuses on random memory locations that are potentially vulnerable.

\subsection{Laser Fault Attack}
Laser fault injection (Laser FI) is a powerful yet costly fault injection technique that uses ionizing radiation to flip circuit logic. Imagine creating an intentional cosmic ray. This is laser fault injection. The Laser Fault Attack was first introduced in 2002~\cite{skorobogatov2003optical}, inspired by earlier experiments in ionizing radiation on semiconductors~\cite{habing1965use}. 

\begin{figure}
    \centering
    \includegraphics[width=1.0\linewidth]{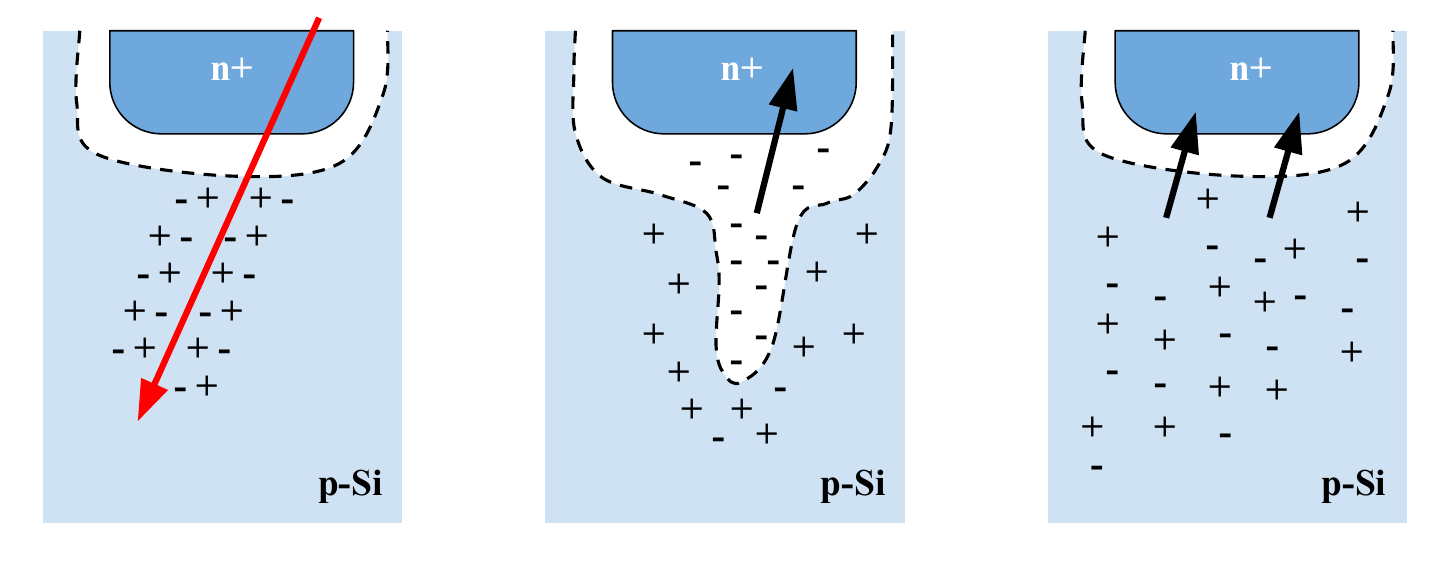}
    \caption{Physical effect of Laser Fault Injection in~\cite{richter2022revisiting} }
    \label{fig:richter2022}
\end{figure}

One significant advantage of laser fault injection over other methods is its ability to inject faults into a limited area. Voltage and clock glitch methods often affect entire systems, whereas a laser targets a specific, small area, allowing for precise control over the fault injection. Photoelectrical Laser Stimulation (PLS) generates photoelectric currents in specific cells~\cite{sarafianos2013electrical}. As shown in Figure~\ref{fig:richter2022}, this causes a reduction in voltage, which changes values~\cite{richter2022revisiting}.

Despite the shrinking node sizes of Moore's Law, numerous studies have demonstrated the continued feasibility of injecting single-bit or single-byte faults into modern ICs with Laser FI. Experiments in 2010 on an 8-bit microcontroller with a 0.35 µm technology node show that it remains practical to perform single-bit or single-byte faults~\cite{agoyan2010single}. Laser FI remains feasible and reproducible with careful beam tuning and timing~\cite{agoyan2010flip}. Later experiments in 2016 inject into 45~nm technology nodes, concluding that individual bit faults are feasible for both 90~nm and 45~nm chips, although the latter presents certain limitations~\cite{17selmke2016precise}. Additional studies have further validated the feasibility of single-bit fault injection using laser fault attacks~\cite{18dutertre2018laser, schellenberg2016large}.

\subsubsection{Controllable Laser Fault Injection}
\begin{figure}
    \centering
    \includegraphics[width=1.0\linewidth]{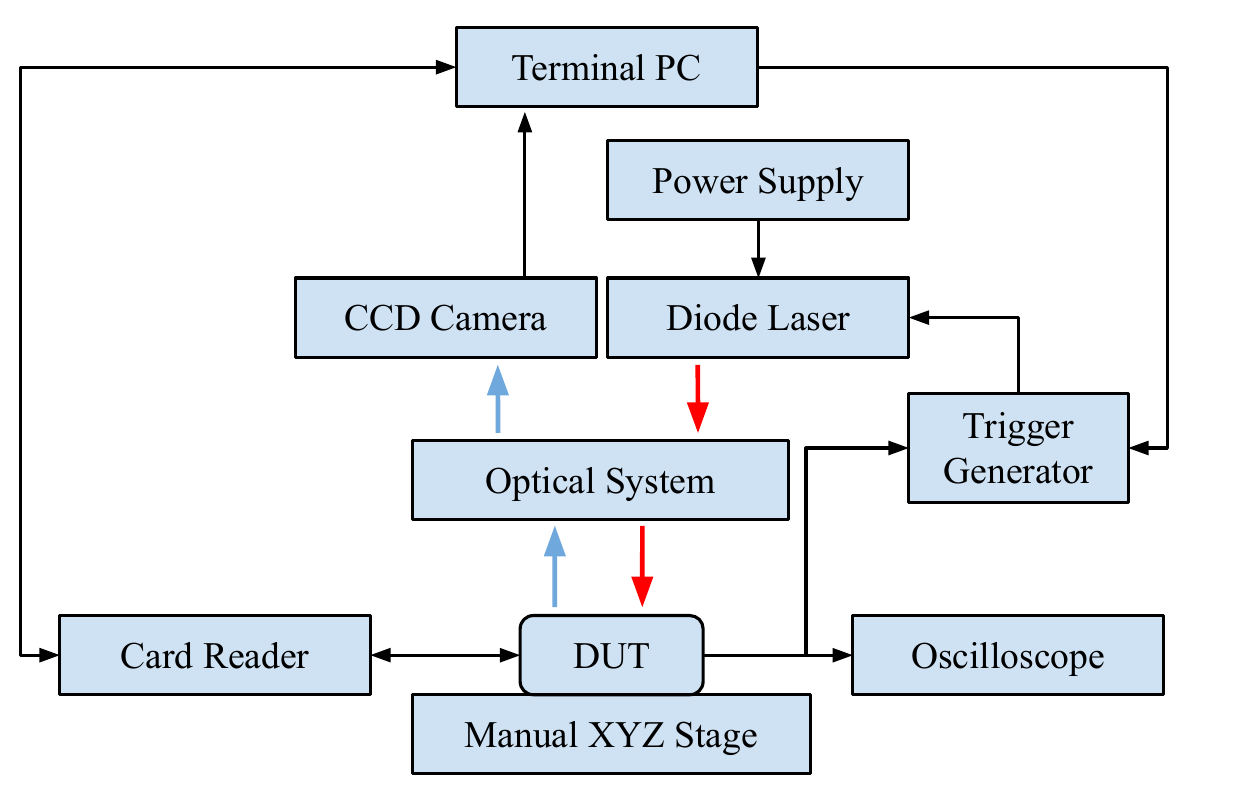}
    \caption{Diagram of the Laser Setup in~\cite{sakamoto2020laser} }
    \label{fig:sakamoto2020}
\end{figure}

How are laser-injected faults fully controlled? Specialized lasers.
Fine-tuning the parameters of laser fault injection results in perfect faults: able to switch any register value between `0' and `1' by adjusting the location~\cite{courbon2014adjusting}. Experiments in 2020, as shown in Figure~\ref{fig:sakamoto2020}, demonstrate that laser faults enable controllable instruction replacement, circumventing existing countermeasures~\cite{sakamoto2020laser}.

\textbf{Considerations:}
It requires a decapsulated IC and removing the protective covering of an IC is an expensive, labor intensive board modification. It also has limitations based on node size. It is extremely precise.

\subsubsection{Low-Cost Laser Fault Injection}

The first laser fault injection used an inexpensive \$30 second-hand flashgun and an \$8 laser pointer to cause random faults~\cite{skorobogatov2003optical}. On the other side of the spectrum, a million dollar nano-focused X-ray beam was used to target a single transistor~\cite{anceau2017nanoxray}. This has motivated research into high-precision and low-cost laser fault injection~\cite{kelly2020high}.

\textbf{Considerations:}
It is still more expensive compared to the previously discussed techniques. It is becoming more affordable.

\begin{figure}
    \centering
    \includegraphics[width=1.0\linewidth]{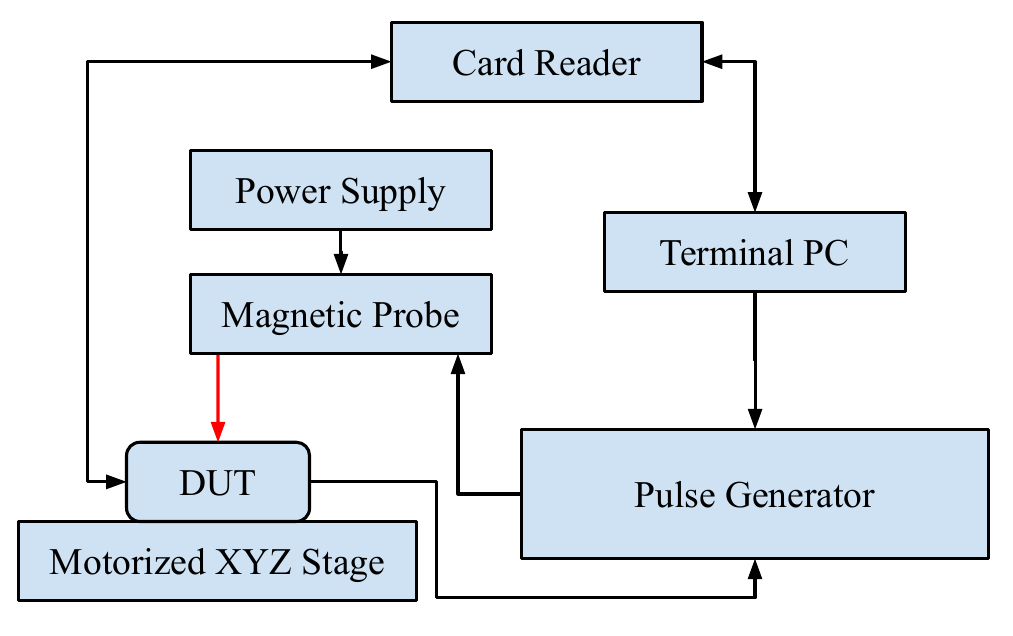}
    \caption{EM Pulse Injection Bench in~\cite{dehbaoui2012injection} }
    \label{fig:dehbaoui2012}
\end{figure}

\subsection{Electromagnetic Fault Injection}

Electromagnetic fault injection (EMFI), as shown in Figure~\ref{fig:dehbaoui2012},  involves blasting an IC with a brief burst of electromagnetic energy from closely placed magnetic coils (Magnetic Probe)~\cite{dehbaoui2012injection}. Imagine sending a wave of electrons at an electronic device. If it is not destroyed, this is EMFI. The Electromagnetic Pulse (EMP) Fault Attack was first introduced in 2012~\cite{dehbaoui2012injection}. The resulting EM disturbance generates a highly localized transient voltage within the IC, altering the circuit's behavior~\cite{dehbaoui2012injection}. Subsequent work further confirmed the locality of EM-generated faults and demonstrated the ability to induce both bit-set and bit-reset faults~\cite{ordas2015evidence}. Unlike Laser FI, many studies have successfully designed low-cost EM probes for EMFI attacks~\cite{cui2017badfet,abdellatif2020silicontoaster, delarea2022practical}.

\textbf{Considerations:}
It does not require board modification, unlike Laser FI. It is cheaper. It is also very precise, but it will not target a single bit or transistor. It has a local effect.

\begin{figure*}
    \centering
    \includegraphics[width=7in]{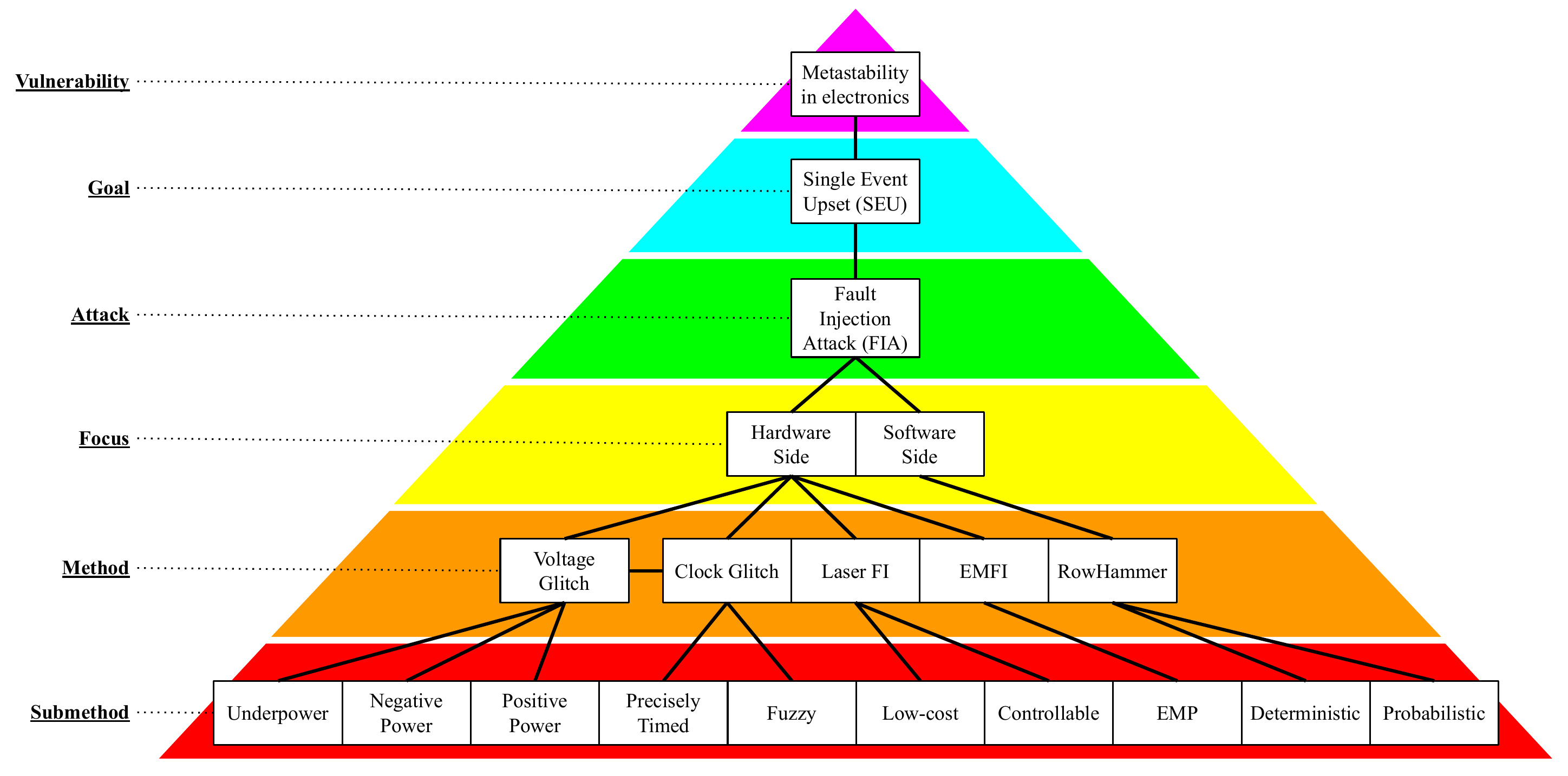}
    \caption{Taxonomy of Fault Injection Techniques}
    \label{fig:taxonomy}
\end{figure*}

\subsection{Taxonomy Chart}
After performing a systematic review of existing work on
FIA attacks, we constructed a taxonomy chart. See Figure~\ref{fig:taxonomy}.

\section{Current State-of-the-Art}
\label{s:current}

The section provides the estimated costs and performance of each method. We derive numbers from the retail prices of publicly available tools and from the reports of existing experiments. These numbers are only estimates.

\subsection{Estimated Costs}

\begin{table}
    \caption{Estimated Costs Table}
    \label{ECT}
    \begin{tabular}{| m{2cm} |  m{2.5cm} |  m{2.5cm} |}
    \hline
    \textbf{FIA Technique}  & \textbf{Low Cost}  & \textbf{High Cost}\\
    \hline
    Clock/Voltage Glitching & 50 USD~\cite{picoglitcher} & 600 USD~\cite{chipwhisperer} \\
    \hline
    EMFI & 4,000 USD~\cite{chipshouter}  & 10,000+ USD~\cite{debunkingMyths} \\
    \hline
    LASER (Optical) & 500 USD~\cite{cheaplaser}  & 50,000+ USD~\cite{howpractical} \\
    \hline
    RowHammer & N\slash A  & N\slash A \\
    \hline
    \end{tabular}
\end{table}
    
  
    
    
    
    
As shown in Table~\ref{ECT}, the order of FIA techniques from lowest to highest cost is as follows: Clock/Voltage Glitching, EMFI, and Laser FI. Refer to Appendix~\ref{SecondAppendix} for more examples of costs.

\subsection{Estimated Performance}

\begin{table}
    \caption{Estimated Performance Table}
    \label{EPT}
    \begin{tabular}{| m{2cm} |  m{5.5cm} |}
    \hline
    \textbf{FIA Technique}  & \textbf{Success Rate}\\
    \hline
    Clock/Voltage Glitching & 1.4\%~\cite{MLclockglitch} \\
    \hline
    EMFI & $\sim$2\%~\cite{espressifEMFI} \\
    \hline
    LASER (Optical) & $\sim$100\%~\cite{stateoftheartLaser} \\
    \hline
    RowHammer & N\slash A  \\
    \hline
    \end{tabular}
\end{table}

As shown in Table~\ref{EPT}, the order of FIA techniques from highest to lowest performance is as follows: Laser FI, EMFI, and Clock/Voltage Glitching.
    

\subsection{Real-life Examples of Hacks}

Examples of hacks are not endorsements. We do not condone unauthorized FIA.

\begin{itemize}[label={}]
    \item \textbf{Clock/Voltage Glitching}:
        \begin{itemize}
            \item Smart Meter Device~\cite{recessimSMD}
            \item SpaceX Starlink Terminal~\cite{blackhatSX}
            \item Nintendo Switch Mod~\cite{nintendoSwitchMod}
        \end{itemize}
    \item \textbf{EMFI}:
        \begin{itemize}
            \item Automotive Electronic Control Units~\cite{hardwearioNL}
            \item Other Micro-controller Units (MCU)~\cite{espressifEMFI}
        \end{itemize}
    \item \textbf{LASER (Optical)}:
        \begin{itemize}
            \item Crypto Hardware Wallets~\cite{hardwearioNLcrypto}
        \end{itemize}
    \item \textbf{RowHammer}:
        \begin{itemize}
            \item Firefox Javascript Exploit~\cite{rowhammerJS}
        \end{itemize}
\end{itemize}

\subsection{Observed Trends}

The current research focus is on optimizing both the software and hardware for finding exploitable glitches. There are machine learning tools to assist in the exhaustive process of finding clock glitching parameters~\cite{MLclockglitch}. There are temperature-controlled automated EM probes with genetic search algorithms to assist in optimal placement~\cite{hardwearioNL}.

\section{Measuring Effectiveness}
\label{s:effective}
When studying cyber security, we separate research into two fields. Vulnerability Detection and Vulnerability Mitigation. This is also known as ``Red Team" and ``Blue Team". Since this is a beginner-friendly introduction to Fault Injection Attacks, we will be focusing mainly on vulnerability detection. This will help demonstrate how critical and emerging FIAs are as a threat.

This section provides the essential knowledge for understanding vulnerability detection. After we cover what is being measured, we will examine and replicate an existing vulnerability detection tool. Exploitable code bugs are found using Program Analysis Techniques such as Symbolic Execution and Fuzzing. Different fault models for detecting FIAs vary based on abstraction level and complexity.

\subsection{Symbolic Execution}
\begin{figure}
    \centering
    \includegraphics[width=1.0\linewidth]{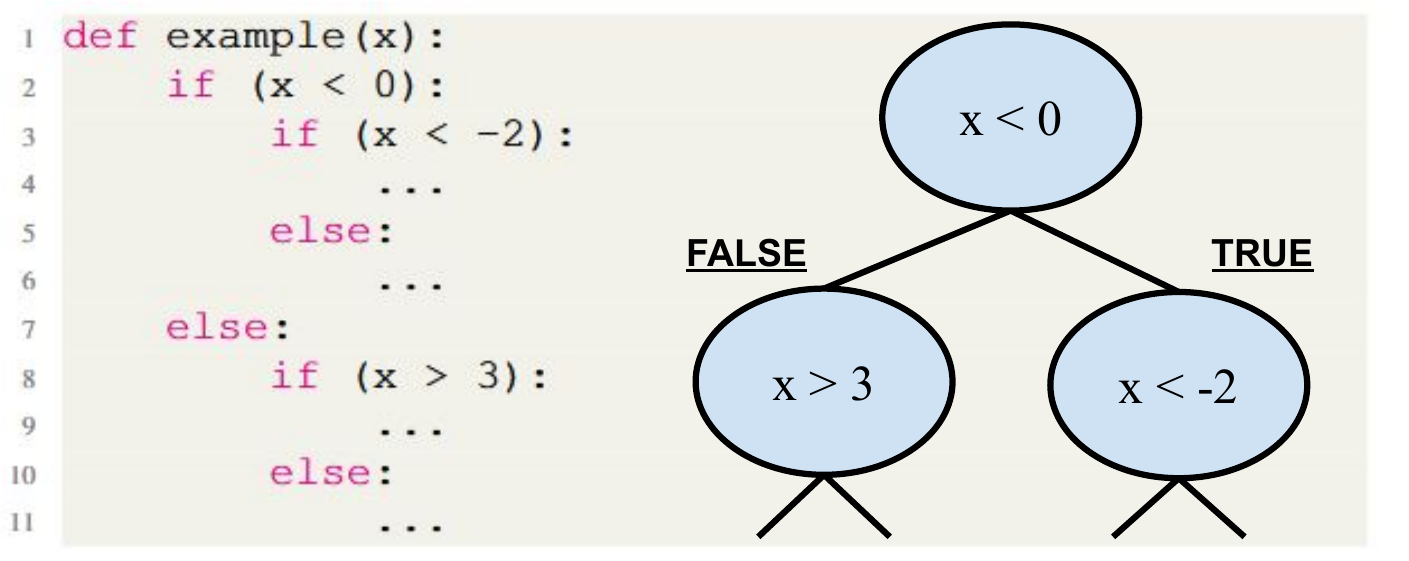}
    \caption{Basic Symbolic Execution}
    \label{fig:symbolicexec}
\end{figure}
Symbolic execution is a software testing technique widely used for vulnerability detection. It was introduced in the 1970s as a method for evaluating and testing programs~\cite{boyer1975select, king1976symbolic, howden1976experiments, clarke1976program}. Compared to traditional fuzzing, which uses brute-force inputs, symbolic execution explores multiple execution paths with variable inputs~\cite{baldoni2018survey}.

We choose to focus on symbolic execution to analyze the propagation of faults due to its unique ability to explore all potential outcomes (Figure~\ref{fig:symbolicexec}). However, this creates a large search space, resulting in long testing times. \textit{Future research} is needed to reduce this search space by finding additional relevant constraints.

\subsection{Fault Modeling}

A fault primitive (FP) is a mathematical notation used to model the behavior of vulnerabilities. In traditional software memory corruption, the exploitable primitive is generalized as access to an out-of-bounds or dangling pointer~\cite{szekeres2013sok}. This access enables the unintended reading or writing of code. A fault primitive therefore represents the ``attack surface" for fault injection.
\begin{figure}
    \centering
    \includegraphics[width=1.0\linewidth]{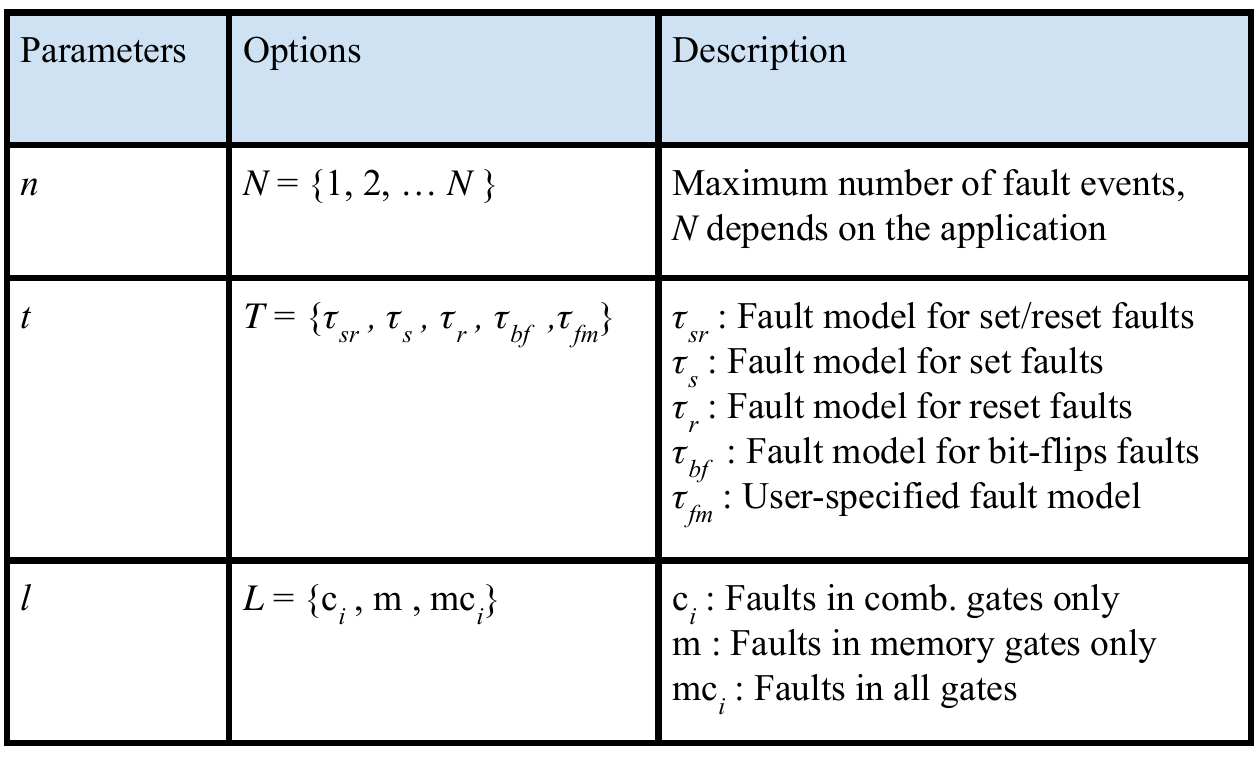}
    \caption{Parameters to Accurately Model Fault Injections in~\cite{richter2022revisiting} }
    \label{fig:richter2022types}
\end{figure}
From the mathematical model of faults, there are five types of faults (Figure~\ref{fig:richter2022types}). Since faults cause the alteration of a single bit or multiple bits of an instruction or data, one or more combinations of fault models are used with varying parameters to generalize the model.

Various primitive fault models have been proposed~\cite{balasch2011depth, nabhan2023tale, olivier2022triple, trouchkine2021fault, khuat2021laser}. These fault models describe the behavior of a fault as a single bit, byte, or instruction corruption.

\begin{figure}
    \centering
    \includegraphics[width=1.0\linewidth]{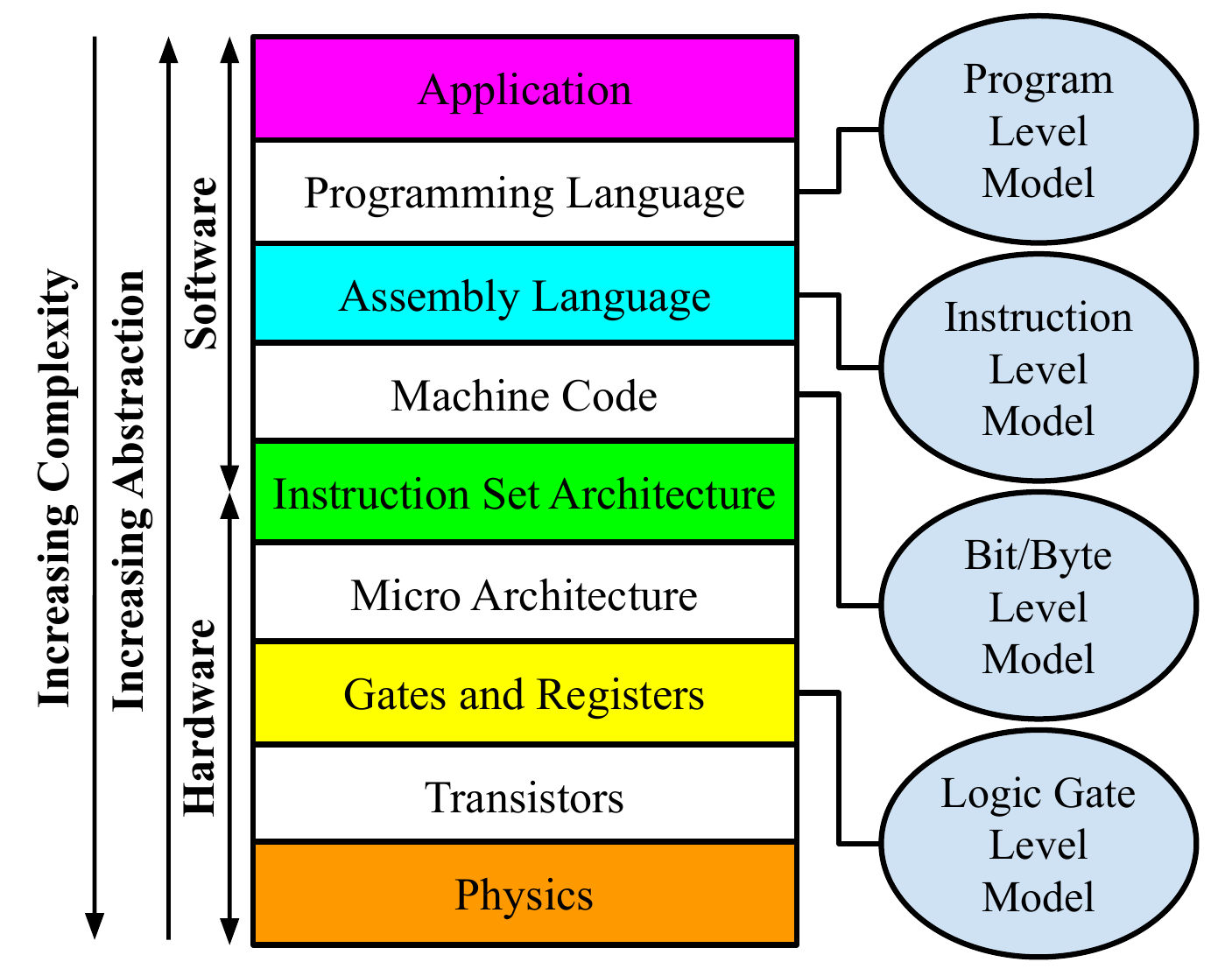}
    \caption{Layers of Abstraction and Fault Model Level}
    \label{fig:modellevels}
\end{figure}

At the lowest level, a fault injection is modeled as a Single Event Upset and Electrical Transient on a single bit. Here are the three basic fault models for a single bit.

\begin{description}
    \item \textbf{Set fault}
    Changes the target bit to a `1'
    \item \textbf{Reset fault}
    Changes the target bit to a `0'
    \item \textbf{Bit-flip fault}
    Changes the target bit to its opposite.
\end{description}

At a higher level (Figure~\ref{fig:modellevels}), corrupted bits or bytes in the binary are interpreted as alterations to opcodes and operands in the machine code. This is then interpreted as alterations to assembly instructions as the corruption of data in a LOAD or STORE instruction or the unintended skip of an instruction with a NOP. This is then finally propagated to the Program Source Code Level as control-flow hijacking or data-integrity corruption.

However, for a fault to be exploitable, it needs to pass on errors from the hardware side to the software side. This raises the question: \textbf{How effective are existing fault injection vulnerability detection tools?}

\subsection{Limitations Problem}

\begin{figure}
    \centering
    \includegraphics[width=1.0\linewidth]{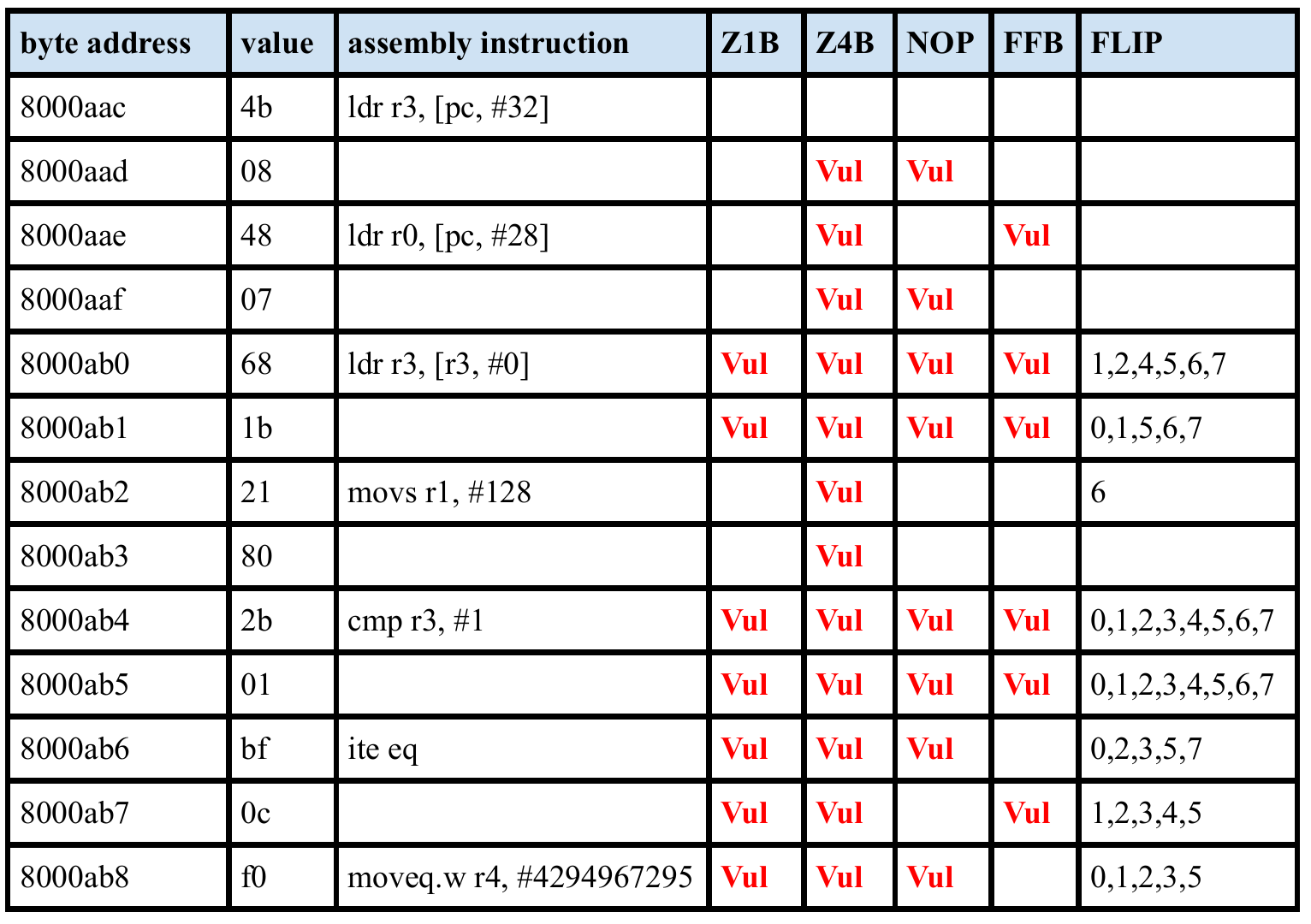}
    \caption{Detected Vulnerable Instructions in~\cite{given2020combined}}
    \label{fig:givenflagged}
\end{figure}

Fault models using simple fault primitives have not yielded sufficient results (Figure~\ref{fig:givenflagged}), flagging almost every line of code as vulnerable~\cite{given2020combined}. If everything is labeled vulnerable, the tool does not provide useful information. For example, in a bit-flip model, nearly every bit or byte in binary code could be a target. This highlights a need for relevant modeling constraints that reduce the problem space to what is physically or logically useful.
\textit{Future research} is needed to create realistic fault models that better capture the behavior between the hardware side and the software side, addressing the limitations of existing models.


\section{Replication Insights}
\label{s:insight}

This section provides new insights in the form of a proposed synthesis model. We provide an outline of our replication study and our examination of an existing vulnerability detection tool. This identifies a research focus for \textit{future work}.

\subsection{Hypothesis}
After observing current trends, we hypothesize that fault modeling could be improved by taking a more holistic approach. Unlike traditional software vulnerabilities, the nature of FIAs involves the integration of two distinct aspects: the hardware fault induction process and the software fault exploitation. Due to this dual nature, research on FIAs has been divided into two disconnected fields: one focused on the physical factors that induce hardware faults and the other on the software-side  detection using fault primitives. 

Several attempts have been made to detect or harden programs against FIA vulnerabilities at the source code level~\cite{25potet2014lazart, reichling2022faulthunter, boespflug2023compositional, 26lacombe2024combining, witteman2008secure}. To our knowledge, existing methods have not adequately addressed the levels below the assembly code. 
We propose a synthesis model that finds exploitable code and assembly instructions, but with awareness of what is feasible in machine code.

\begin{figure}
    \caption{Verify.c}
    \label{verify_c}
    \begin{lstlisting}[language=C++]
#define SIZE_OF_PIN 4
#define maxTries 3
typedef unsigned char BYTE;
BYTE triesLeft = maxTries;
BYTE authenticated = 0;
BYTE pin[4] = {(char)1, (char)2, (char)3, (char)4};

BYTE Verify(char buffer[4]) {
BYTE i;
// BLOCK entry
// No comparison if PIN is blocked
if(triesLeft <= 0) goto FAILURE;
// BLOCKS bb (initialisation), 
//bb4 (loop condition)
    // Main Comparison
    for(i = 0; i < 4; i++)   // BLOCK bb3 (i++)
// BLOCK bb1
        if(buffer[i] != pin[i]) {
// BLOCK bb2
            triesLeft--;
            authenticated = 0;
            goto FAILURE;
        }
// BLOCK bb5
        // Comparison is successful
        triesLeft = maxTries;
        authenticated = 1;
        return EXIT_SUCCESS;
// BLOCK FAILURE
    FAILURE : return EXIT_FAIULRE;
// BLOCKS bb6, return: exit with
// the return value
}
    \end{lstlisting}
\end{figure}

\subsection{Methodology}
Inspired by the consolidated fault injection adversary model proposed in 2022~\cite{richter2022revisiting}, we design our model by combining a ``Static Analysis and Dynamic Symbolic Execution" approach~\cite{26lacombe2024combining} with a ``High-Level and Low-Level" approach~\cite{30riviere2014combining} for detecting fault injection vulnerabilities.

We then compare our proposed model with Lazart~\footnote{https://gricad-gitlab.univ-grenoble-alpes.fr/securitytools/lazart}, an existing tool from 2014~\cite{25potet2014lazart}. Lazart uses symbolic execution to find FIA vulnerabilities that have the potential to alter control flow. Lazart operates at the intermediate level, between source code and assembly instructions, using Low Level Virtual Machine Intermediate Representation (LLVM-IR) code for analysis. 

We use a code snippet from a PIN verification algorithm for analysis. Figure~\ref{verify_c} shows the source code.

Afterwards, we perform a literature review to compare the various hardware and software vulnerability detection methods with varying fault models.

Since this is intended to be a beginner-friendly introduction, we will be providing only preliminary evidence and insights for \textit{future research}. An in-depth concrete comparison of every FIA vulnerability detection tool is outside the scope of this paper. 

\subsection{Technique Detail}

We assume that Cortex-M3 is the target device. We assume that ARM v7-M is the target instruction set architecture (ISA) when analyzing the instruction assembly and machine code.

\section{Evaluation}
\label{s:evaluation}

This section provides preliminary evidence for our proposed synthesis model by replicating a previous solution and comparing our proposed solution. We also compare multiple vulnerability detection methods and fault models by reviewing the literature.  

\subsection{Detection Comparison}

Lazart works by producing a control flow graph (CFG) based on the source code. The CFG is then color coded to identify vulnerable code blocks. Vulnerable code blocks are defined as those that reach a critical goal if mutated. Lazart then generates mutated code to show the resulting control flow changes of potential fault injections.

However, Lazart suffers from both false positives and false negatives. It uses a detection fault model that fails to integrate the hardware side of FIA. We will now use our fault model to examine the mutants generated by the Lazart. We will then clarify that some vulnerabilities identified by the Lazart are not feasible in real-world scenarios.

In contrast, our model focuses on the assembly and machine code level of the program. Due to this level of granularity, our model eliminates false-negative results that arise at the source code or intermediate level.

Our fault model discovers a possible attack scenario in a location overlooked by Lazart. Based on our model, faulting the FAILURE block without hijacking any control flow is also possible. This poses a security vulnerability to the code by corrupting the data of the return value. Figure~\ref{verify_s} shows the assembly of the FAILURE block. The assembly instruction moves a 1 to the return register. Our fault model identifies this register as a potential target for fault injection.

In terms of feasibility, for a fault injection to hijack control flow, the Hamming distance between the desired destination and the original destination needs to be only 1 bit or 1 byte. Since fault injection works by flipping or setting only 1 bit, there is a real-world constraint imposed by the hardware architecture. 

We use Verify2 to demonstrate a case where Lazart misclassifies a line of safe code as vulnerable. Figure~\ref{verify2_c} is the implementation of Verify2. The Verify2 code has the same functionality as the original Verify. The only difference is additional code that protects the return value. Here, 0xAA will be returned if correct and 0X55 if not. We choose these values to create a Hamming distance that is not feasible for fault injection, where the stored binary representations are significantly different.  

Based on the results from Lazart, there is a Fault Injection vulnerability at the stored return variable. After hardening the code, Lazart still detects this as a valid attack due to a lack of awareness of lower-level machine code.

Overall, Lazart highlights most parts of programs as vulnerable to FIAs. This substantiates the identified problem that we discussed
in Section 5.3

\begin{figure}
\caption{Veify2.c, Verify.c hardened}
\label{verify2_c}
\begin{lstlisting}[language=C++]
#define SIZE_OF_PIN 4
#define maxTries 3
#define EXIT_SUCCESS_BYTE 0xAA
#define EXIT_SUCCESS_FAILURE 0X55
typedef unsigned char BYTE;
BYTE triesLeft = maxTries;
BYTE authenticated = 0;
BYTE pin[4] = {(char)1, (char)2, (char)3, (char)4};

BYTE Verify2(char buffer[4]) {
    BYTE i;
// BLOCK entry
    // No comparison if PIN is blocked
    if(triesLeft <= 0) goto FAILURE;
// BLOCKS bb (initialisation), 
//bb4 (loop condition)
    // Main Comparison
    for(i = 0; i < 4; i++)   // BLOCK bb3 (i++)
// BLOCK bb1
        if(buffer[i] != pin[i]) {
// BLOCK bb2
            triesLeft--;
            authenticated = 0;
            goto FAILURE;
        }
// BLOCK bb5
        // Comparison is successful
        triesLeft = maxTries;
        authenticated = 1;
        return EXIT_SUCCESS_BYTE
// BLOCK FAILURE
    FAILURE : return EXIT_FAIULRE_BYTE;
// BLOCKS bb6, return: exit with
// the return value
}
\end{lstlisting}
\end{figure}

\begin{figure}
\caption{FAILURE block in Verify.s}
\label{verify_s}
\begin{lstlisting}[language={[x86masm]Assembler}]
Verify:
@ main.c:25:     FAILURE: return EXIT_FAILURE;
        movs    r3, #1  @ _12,
\end{lstlisting}
\end{figure}

\subsection{Literature Review}
After performing a literature review, we constructed a table to compare the various methods of finding hardware and software FIA vulnerabilities. See Table~\ref{FIA_V_A}.

\begin{table*}[t!]
        \caption{FIA Vulnerability Detection Technique Analysis}
        \label{FIA_V_A}
        \begin{tabular}{| m{6cm} |  m{2cm} |  m{2.5cm} |  m{4cm} |  m{1cm} |}
        \hline
        \textbf{Method} & \textbf{Target(HW/SW)}  & \textbf{Fault Model Level}  &  \textbf{Fault Model Detail}  & \textbf{Density}    \\
        \hline
        \textit{Combining static analysis and dynamic symbolic execution in a toolchain to detect fault injection vulnerabilities}~\cite{26lacombe2024combining}  & Software  & Program level  & Control-flow fault; Data fault  & High     \\
        \hline
        \textit{Fault attack vulnerability assessment of binary code}~\cite{27brejon2019fault}   & Software  & Program level  & Control-flow fault; Data fault  & High    \\
        \hline
        \textit{Minotaur: Adapting software testing techniques for hardware errors}~\cite{28mahmoud2019minotaur}  & Software  & Instruction level  & Data fault  & High    \\
        \hline
        \textit{Fiver--robust verification of countermeasures against fault injections}~\cite{21richter2021fiver}  & Hardware  & Gate level  & Bit-level faults  & High     \\
        \hline
        \textit{An automated and scalable formal process for detecting fault injection vulnerabilities in binaries}~\cite{29given2017automated}  & Software  & Binary level  & Byte level faults  & High     \\
        \hline
        \textit{Combining high-level and low-level approaches to evaluate software implementations robustness against multiple fault injection attacks}~\cite{30riviere2014combining}  & Software  & Program level  & Instruction skip  & High    \\
        \hline
        \textit{Lazart: A symbolic approach for evaluation the robustness of secured codes against control flow injections}~\cite{25potet2014lazart}  & Software  & Program level  & Control-flow faults; Data faults  & High \\
        \hline
        \textit{Combined software and hardware fault injection vulnerability detection}~\cite{given2020combined} & Both & Instruction level & Byte level faults & High  \\
        \hline
        \textit{Cryptographic fault diagnosis using VerFI}~\cite{22arribas2020cryptographic} & Hardware & Instruction level & Gate level faults & Low \\
        \hline
        \textit{SymPLFIED: Symbolic program-level fault injection and error detection framework}~\cite{pattabiraman2008symplfied} & Software & Program level & Data fault & High \\
        \hline
        \textit{Sofi: Security property-driven vulnerability assessments of ics against fault-injection attacks}~\cite{23wang2021sofi} & Hardware & Gate level & Bit level fault & High \\
        \hline
        \textit{Error analysis and detection procedures for a hardware implementation of the advanced encryption standard}~\cite{24bertoni2003error} & Hardware & Gate level & Bit level fault & Low \\
        \hline
        \textit{An automated and scalable formal process for detecting fault injection vulnerabilities in binaries}~\cite{given2019automated} & Software & Binary level & Bit level fault & High \\
        \hline
        \end{tabular}
    \end{table*}

We found that the various vulnerability analysis methods suffered from a high detection density. In other words, more than a reasonable number of vulnerable spots will be flagged from the source code with all different kinds of fault models. This further substantiates the limitations discussed in Section 5.3.

In comparison to other methods, we propose a fault model that provides program source code level detection with instruction level and binary bit/byte level awareness. We then directly compare our approach with an existing program source code level detection tool with intermediate level awareness. Other experiments, for example, focused mainly on the instruction level or the binary bit/byte level.

\subsubsection{Instruction Experiment}

SymPLFIED is another program-level FIA vulnerability detection tool that was introduced in 2008~\cite{pattabiraman2008symplfied}. However, instead of taking the source code, SymPLFIED takes the assembly instructions as input and analyzes the effect of fault injection on the program with symbolic execution. 

The goal of the assembly language input is to avoid state explosion in the problem space. However, the final result presented by SymPLFIED fails to support this claim. 

SymPLFIED is not able to detect when an instruction is corrupted or replaced. The experiment only examines FIA against registers. Furthermore, no hardware-based experiment is done to prove the correctness of the detection result.

\subsubsection{Binary Experiment}
Another technique detects fault injection vulnerabilities in binaries. Proposed by Given-Wilson et al.~\cite{29given2017automated, given2019automated, given2020combined}, this involves generating mutant binaries based on the input binaries. These mutant binaries are then subjected to model and property checks to determine if vulnerabilities exist.

This detection technique employs six different fault models. The \textbf{FLP} fault model simulates the flipping of a single bit. The \textbf{Z1B} fault model simulates setting a single byte to zero. The \textbf{Z1W} fault model simulates setting a single word to zero. The \textbf{NOP} fault model simulates replacing a single instruction with a NOP instruction. The \textbf{JMP} fault model and the \textbf{JBE} fault model simulate faults in the target address of a jump instruction.

In 2020, Given-Wilson et al.~\cite{given2020combined} modified binary code to find possible fault injections that would achieve an attacker's goal. Afterwards, a hardware experiment was conducted to verify the software detection results.

Although the byte-level fault model used here yields no false positives, in the final report, the coverage of code marked as vulnerable is far too large to provide useful information for hardening. This demonstrates how state-of-the-art detection techniques are not practical in real world cases. For programs with thousands or even millions of lines of code, nearly every line is marked as vulnerable to FIAs, providing little useful insight for programmers looking to improve their code.

\section{Countermeasures}
\label{s:defense}
Since this is a beginner-friendly introduction that focuses on vulnerability detection, we will keep our discussion of vulnerability mitigation short. This will help demonstrate the difficulty of implementing countermeasures against FIAs.

This section provides the essential knowledge for understanding vulnerability mitigation. Dedicated hardware and software are used to defend devices against FIAs.

\subsection{Threat Model}
The goal of a Fault Injection Attack is to bypass software security by inducing unintended hardware behavior. This breaks the assumption that hardware will guarantee proper execution of software, leading to unintended code values and unintended code executions.

\begin{figure}
    \centering
    \includegraphics[width=1.0\linewidth]{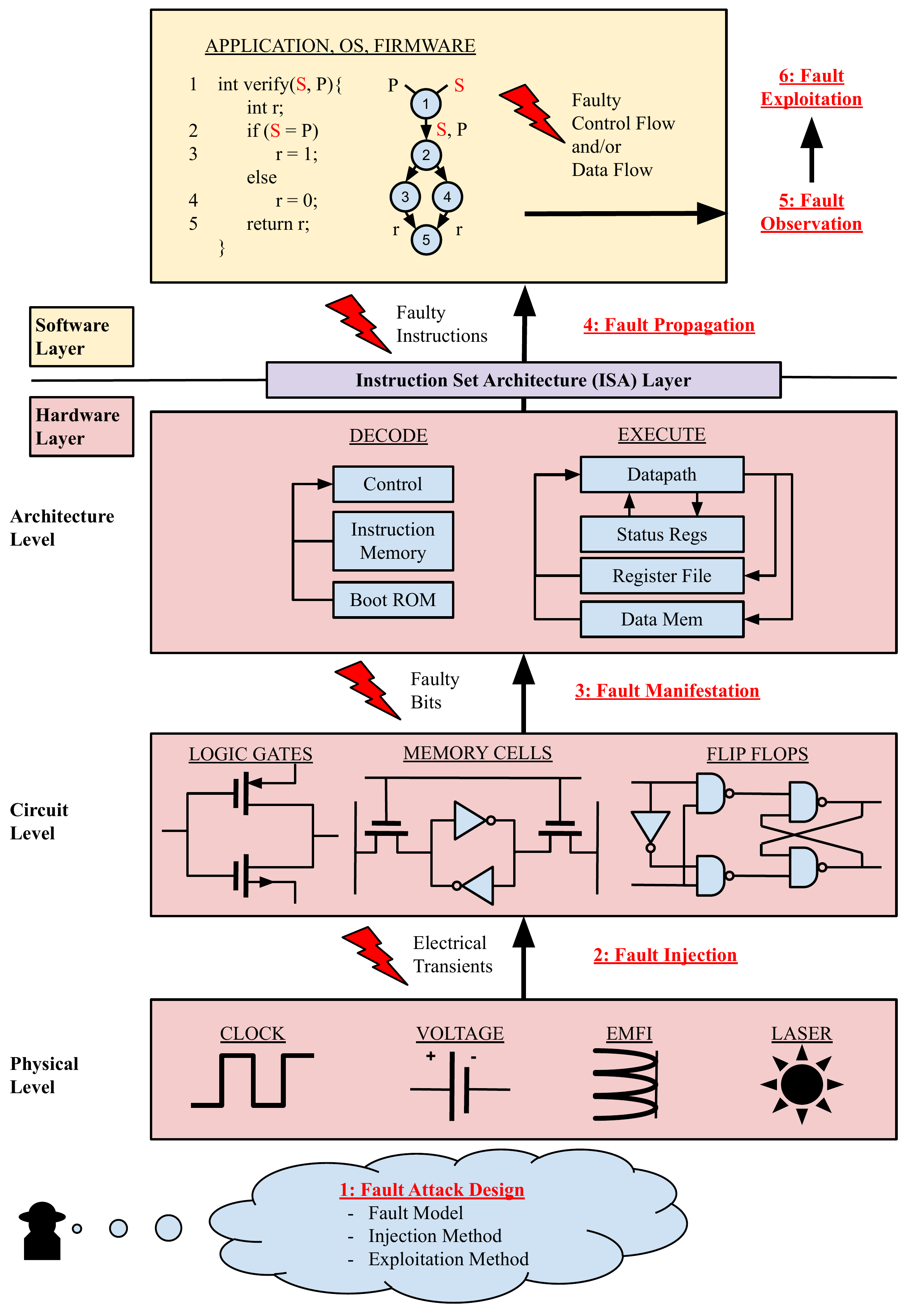}
    \caption{Anatomy of a Fault Injection Attack in ~\cite{yuce2018secure}}
    \label{fig:antatomyfia}
\end{figure}

As shown in Figure~\ref{fig:antatomyfia}, Fault Injection Attacks can be separated into six stages.

\begin{description}
    \item[Fault Modeling]
        The planning stage for the attack
    \item[Fault Injection]
        The method of stressing the hardware
    \item[Fault Manifestation]
        The errors in circuit logic
    \item[Fault Propagation]
        The errors in software logic
    \item[Fault Observation]
        The measuring of changes in behavior
    \item[Fault Exploitation]
        The insights or attacks enabled
\end{description}

This is the ``cyber kill chain" of a Fault Injection Attack. In the context of countermeasures, the goal is therefore to detect, prevent, and mitigate any of these stages.

The target of Fault Injection is the hardware layer. The target of Fault Exploitation is the software layer. Devices are therefore defended with both hardware hardening and software countermeasures.

Unlike traditional attacks, Fault Injection Attacks do not require the presence of an existing software bug. This is because Fault Injection can dynamically induce software bugs~\cite{yuce2018secure}. Software countermeasures themselves can also be the target of a Fault Injection Attack~\cite{thibault2022counter}.

Fault Injection can be used to achieve at least three goals.
\begin{description}
    \item[Cryptanalysis]
        To extract side channel information~\cite{giraud2004dfa}
    \item[Logical Attacks]
        To hijack the flow of a program~\cite{bouffard2011flow}
    \item[Reverse Engineering]
        To characterize a target~\cite{courbon2015characterize}
\end{description}

Vulnerability mitigation involves increasing the cost and difficulty of achieving these goals~\cite{yuce2018secure}. For example, if an attacker is forced to inject similar faults in a short period of time, the attack is no longer considered viable~\cite{barenghi2010countermeasures}. Jointly optimizing the parameters of multiple glitches or pulses is generally considered too complex~\cite{breakingEspressif}.

\subsection{Software Countermeasures}
The state-of-the-art of software fault countermeasures are divided into two paradigms: Detection-based and Infection-based countermeasures~\cite{barbu2021high}. The first involves redundancy checks and repeating instructions~\cite{barenghi2010countermeasures}. The second involves obfuscating the injected error so that no meaningful information can be recovered from the faulty output~\cite{sungming2002immune}.

Multiple software countermeasures exist against Fault Injection Attacks ~\cite{barenghi2010countermeasures, sungming2002immune, boespflug2020countermeasures, barbu2021high, kiaei2021rewrite, medwed2010continuous}.
For binary code where there is no access to the source code, a method of rewriting binary executables to automatically apply software countermeasures was proposed~\cite{kiaei2021rewrite}. For RowHammer, a method that detects frequently accessed rows and refreshes heavily victimized rows was proposed~\cite{seyedzadeh2016counter}. However, software countermeasures have significant performance costs and can have unwanted side effects~\cite{schneider2016parti}.

\subsection{Hardware Countermeasures}

Similar to software countermeasures, hardware countermeasures are often implemented as spatial, temporal or informational redundancies~\cite{boespflug2020countermeasures}. The most common strategy is reading out sensitive bits multiple times and ignoring potentially faulty values~\cite{barel2006sorcerer, moro2014counter}.

Multiple hardware countermeasures exist against Fault Injection Attacks ~\cite{barel2006sorcerer, moro2014counter, he2016ring, seyedzadeh2016counter, endo2012efficient, schneider2016parti, jiang2022machine}.
For Laser Fault Injection, a logic gate-level countermeasure was proposed that auto-detects a frequency disturbance caused by lasers~\cite{he2016ring}. To prevent side channel attacks and Differential Fault Analysis (DFA), a technique called Wave Dynamic Differential Logic (WDDL) was proposed ~\cite{tiri2004logic, selmane2004WDDL}. Hardening the hardware itself is fundamental because defending with only software is not feasible~\cite{debunkingMyths}.

\section{Misconceptions}
\label{s:misconception}
This section answers common questions about Fault Injection Attacks. This will help prevent any misconceptions or false assumptions.

\subsection{Is physical access required for fault injection?}
No, RowHammer is a purely software attack that does not require physical access~\cite{14kim2014flipping}. In addition, software-exposed energy management systems enable remote voltage glitch attacks~\cite{tang2017CLKSCREW}. Most of the techniques are cyber-physical in nature.

\subsection{Is fault injection effective on chips with high clock frequency?}
Yes, in fact, clock frequency can increase the success rate for both voltage glitching and EMFI~\cite{koffas2022clock}. Clock frequency does not prevent fault injection.

\subsection{Is fault injection effective on multi-core chips?}
Yes, a dual core CPU with the cores verifying each other can help to mitigate Fault Injection Attacks, but not fully prevent them~\cite{wiersma2017dual}. Multiple cores do not prevent fault injection.

\subsection{Is success rate the same as attack feasibility?}
No, feasibility is better described by average time for success~\cite{debunkingMyths}. If one attack method has a success rate of 1\% and can be attempted 10 times per minute, the average time for success is 10 minutes. If another attack method has a success rate of 0.1\% and can be attempted 1000 times per minute, the average time for success is 1 minute. Waiting 1 minute is better than 10, despite the lower success rate.

\section{Conclusion}
\label{s:conclusion}
As an entry-level technique, we recommend starting with Clock/Voltage Glitching. 
For the highest precision, we recommend considering Laser FI. However,
it is also the most expensive and labor intensive technique.
In general, we recommend EMFI as the most versatile, striking a balance between cost and precision. 

There is a lot of work to be done in optimizing both the software and hardware for finding exploitable fault injection glitches. This research focus has been demonstrated by the current state-of-the-art and our replication study.

FIA is a critical and emerging cyber-physical threat with difficult countermeasures. The future of security research is no longer searching for traditional software exploits, but dynamically creating exploits.




\bibliographystyle{IEEEtran}
\bibliography{references}

\clearpage
\appendices

\section{Overview of the Techniques in~\cite{howpractical}}
\label{SecondAppendix}
\noindent\begin{minipage}{\textwidth}
    \centering
    \includegraphics[width=1.0\linewidth]{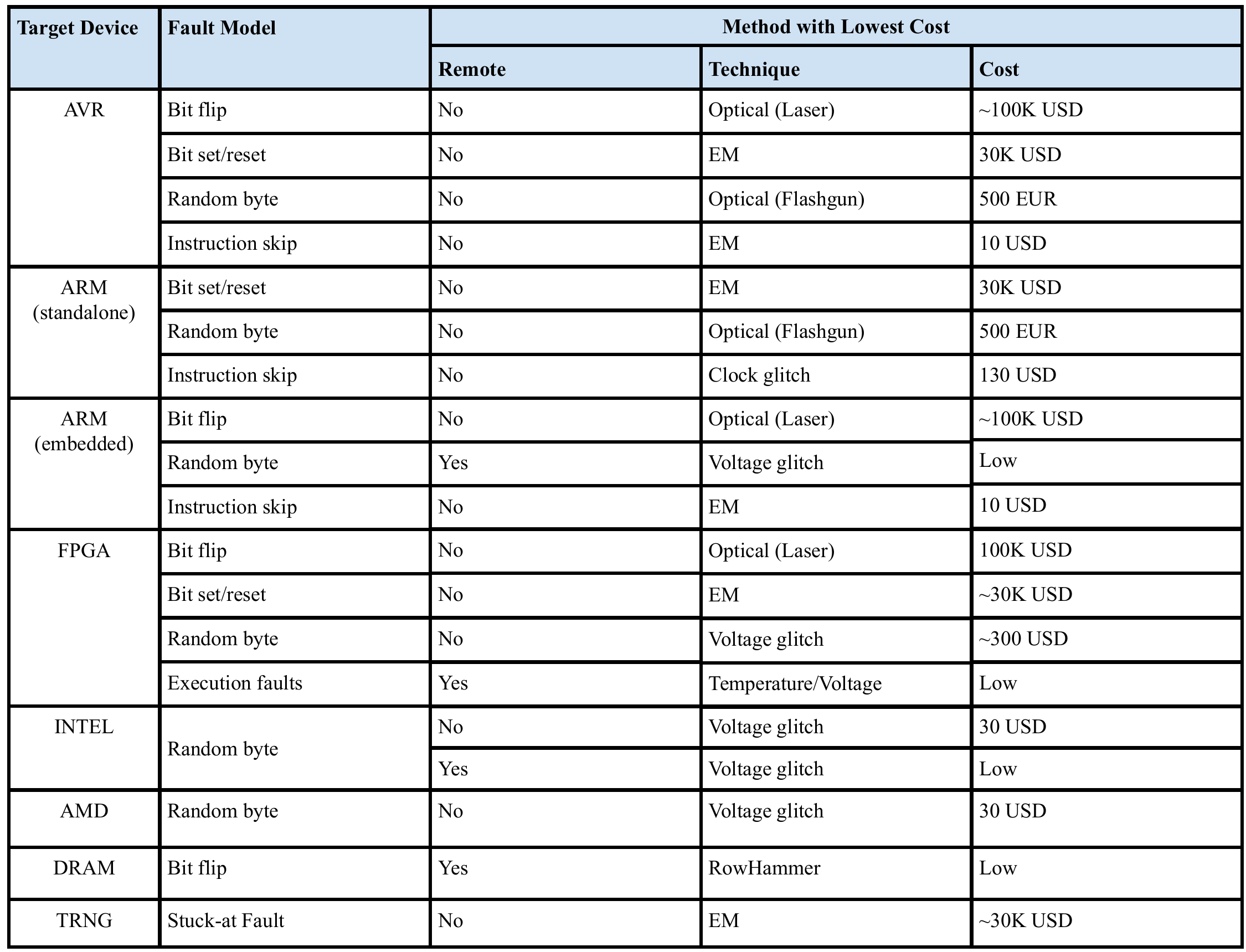}
    \label{fig:howpractical}
\end{minipage}
\clearpage
\section{Timing Constraint as expressed in~\cite{zussa2013power}}
\label{FourthAppendix}
\textbf{Parameters}:
\begin{itemize}
    \item
    \textbf{$\mathrm{T_{clk}}$:}
    The Shared Clock Period
    \item
    \textbf{$\mathrm{D_{pMax}}$:}
    The Max Data Propagation Time (The Most Critical, Longest Path)
    \item 
    \textbf{$\mathrm{T_{clk2q}}$:}
    The Clock Delay for Register
    \item 
    \textbf{$\mathrm{T_{skew}}$:}
    The Phase Difference between Clock Signals of Two Registers 
    \item 
    \textbf{$\mathrm{T_{setup}}$:}
    The Min Time for Stable D flip-flop Input
\end{itemize}

\hfill

\textbf{Equation}:
\begin{equation}
\label{eq:timing}
    T_{clk} > D_{clk2q} + D_{pMax} + T_{setup} - T_{skew}
\end{equation}

\hfill

Attackers induce timing constraint violations by disrupting these parameters. The correctness of the circuit data is critically dependent on timing constraints. It is assumed that the shared clock period $\mathrm{T_{clk}}$ is longer than the most critical (longest) data path through the combinatorial logic. If this assumption is broken, it results in unexpected behavior, leading to a fault injection attack.  

\section{Propagation Time as expressed in~\cite{razavi2021fundamentals}}
\label{FifthAppendix}

\textbf{Parameters}:
\begin{itemize}
    \item
    \textbf{$\mathrm{D_{pMax}}$:}
    The Max Data Propagation Time (The Most Critical, Longest Path)
    \item
    \textbf{$\mathrm{V_{DD}}$:}
    The Power Supply Voltage
    \item
    \textbf{$\mathrm{CL}$:}
    The Load Capacitance
    \item
    \textbf{$\mathrm{V_{th,p}}$:}
    The PMOS Threshold Voltage
    \item
    \textbf{$\mu_p$:}
    The Hole Mobility
    \item
    \textbf{$\mathrm{Cox}$:}
    The Gate Oxide Capacitance
    \item
    \textbf{($\mathrm{W_{p}}$/$\mathrm{L_{p}}$):}
    The Aspect Ratio of the PMOS
\end{itemize}

\hfill

\textbf{Equation}:
\begin{equation}
    D_{pMax} = \frac{C_L \left[ \frac{2|V_{th,p}|}{V_{DD} - |V_{th,p}|} + \ln\left( 3 - \frac{ 4|V_{th,p}|}{V_{DD}} \right) \right]}{\mu_p C_{ox} \frac{W_p}{L_p} \left( V_{DD} - |V_{th,p}| \right)}
\end{equation}

\hfill

This shows that a decrease in the supply voltage leads to an increase in the circuit inverter's propagation time, leading to an overall increase in the propagation time of the combinatorial logic, breaking timing constraints.

\end{document}